\documentclass{mn2e}
\usepackage{epsfig}

\title[Impact cratering and the Oort cloud]{Impact cratering and the Oort cloud}
\author[J.T. Wickramasinghe, W.M. Napier]
{J.T. Wickramasinghe$^1$\thanks{E-mail: janakitara@hotmail.com(JTW);
napierwm@cardiff.ac.uk(WMN)} and W.M. Napier$^1$\\
$^{1}$Cardiff Centre for Astrobiology, Cardiff University, 2 North
Road, Cardiff CF10 3DY, UK }
\usepackage{color}

\begin{document}

\date{Accepted 2008 February 11.  Received 2008 February 9; in original form 2007 December 7}

\pagerange{\pageref{firstpage} -- \pageref{lastpage}} \pubyear{2008}

\maketitle

\label{firstpage}

\begin{abstract}
We calculate the expected flux profile of comets into the planetary
system from the Oort cloud arising from Galactic tides and
encounters with molecular clouds. We find that both periodic and
sporadic bombardment episodes, with amplitudes an order of magnitude
above background, occur on characteristic timescales
$\sim$25--35~Myr. Bombardment episodes occurring preferentially
during spiral arm crossings may be responsible both for mass
extinctions of life and the transfer of viable microorganisms from
the bombarded Earth into the disturbing nebulae. Good agreement is
found between the theoretical expectations and the age distribution
of large, well-dated terrestrial impact craters of the past 250
million years. A weak periodicity of $\sim$36~Myr in the cratering
record is consistent with the Sun's recent passage through the
Galactic plane, and implies a central plane density $\sim$0.15~$\rm
M_\odot\,pc^{-3}$. This leaves little room for a significant dark
matter component in the disc.
\end{abstract}


\section{Introduction}

It has been proposed that disturbances of the Oort cloud due to encounters
with spiral arms and molecular clouds would lead to episodes of bombardment
of comets onto the Earth, and that these episodes would cause geological
disturbances and mass extinctions of life such as the Cretaceous-Tertiary
event of 65~Myr~BP (Napier \& Clube 1979). More recently, it has been pointed
out that such episodes may also provide a mechanism for interstellar panspermia
(Napier 2004, 2007; Wickramasinghe 2007). Microorganisms thrown from the
surface of the Earth by large impacts may be ejected by radiation pressure
into star-forming regions within the passing nebula, and so seed protoplanetary
nebulae within them. The transfer time of microorganisms from Earth to protoplanetary
nebula is comfortably less than the time expected for cosmic rays to sterilise
them (e.g. Horneck et al. 2002, Mileikowsky et al. 2004, Bidle et al. 2007).

A key feature of both the extinction and panspermia processes is the
enhanced terrestrial impact rate occurring during the encounter with
the nebula. In the present paper we examine this quantitatively. We
find that order of magnitude increases in the impact rate of comets
on to the Earth occur during encounters with molecular clouds of
$\rm \ga 5\times 10^4 M_\odot$ and that such encounters take place
quite frequently on geological timescales (Napier 2007). If, within a giant
molecular cloud, there are at least 1.1 exosystems with a receptive
planet and an impact environment permitting the escape of
microbiota, life may propagate throughout the habitable zone of the
Galaxy within the lifetime of the Galactic disc ({\it loc. cit.}). 
In addition to these sporadic episodes, periodic impact surges are 
also expected to occur due to the variable stress exerted on the 
Oort cloud by the vertical galactic tide. The impact cratering 
record is thus predicted to be dominated by surges showing a 
weak periodicity. In the present paper we show that the record 
of well-dated impact craters of the last 250~Myr does indeed have 
this character. Thus the major impactors on Earth are probably 
for the most part comets, coming in as `showers'.

\section {Computational and analytic considerations}

\begin{figure}
\vspace{150pt}
 \caption{Flux evolution for 1000~Myr under the
influence of the vertical Galactic tide, computed by numerical
integration of formulae due to Klacka \& Gajdosik (1994). The flux
is into a heliocentric sphere of radius 2000~AU, and is computed for
50,000 Oort cloud comets with $n(a)\propto a^{-2}$, perihelia $>$10,0000 AU and aphelia $<$50,000 AU. Equilibrium is reached on a timescale of
order 300~Myr.} \label{wicknapfig1}
\end{figure}

\begin{figure}
\vspace{150pt} \caption{Flux evolution for 500~Myr into heliocentric
spheres of radii 16~AU (upper curve) and 8~AU (lower curve) under
the influence of the vertical Galactic tide. 25,000 Oort cloud
comets with $n(a)$ as before and a flat initial eccentricity
distribution. Computed from formulae by Fouchard (2004). Equilibrium
is again reached on a timescale of order 300~Myr.}
\label{wicknapfig2}
\end{figure}

The adopted initial configuration of the Oort cloud, being dictated
by its mode of origin, is model-dependent and uncertain, and the
cloud will evolve from this initial state under the influence of
various perturbers. The strongest perturbers acting on the Oort
cloud as a whole are the vertical Galactic tide (Byl 1986) and
passing molecular clouds (Napier \& Staniucha 1982). An orbit
influenced by this tide conserves its semi-major axis but the
initial eccentricities and inclinations vary cyclically; thus an
initial ensemble will relax towards some statistical equilibrium of
eccentricity and inclination.

The evolution of Oort cloud comets under the influence of the
vertical Galactic tide was first computed, in order to find this
equilibrium distribution. For a differential semimajor axis distribution
$n(a)\propto a^{-\gamma}$, initial values of $\gamma$ in the range
$2\leq\gamma\leq 4$ were adopted for the outer Oort cloud (Bailey 1983; 
Fernandez \& Ip 1987), although probably no single power law fits the entire
range (Emel'yanenko et al. 2007). As a measure of the relaxation time, the
flux of comets entering spheres of various radii was computed. Three independent
approaches were used:

(i) Straightforward numerical integration (4th order Runge-Kutta
with error control using the Fehlberg procedure: see for example
Gerald \& Wheatley 1994).

(ii) Equations given by Klacka \& Gajdosik (1994) were integrated.
These were derived by averaging the tidal force over an orbit, and
reduce the evolution in eccentricity $e$ and argument of perihelion
$\omega$ to two first-order differential equations. In this case an
initially isotropic Oort cloud was adopted, and 50,000 comets were
taken with initial semimajor axis distribution $\gamma = 2$, with perihelia
constrained to have $q>10,000$~AU and aphelia $Q<50,000$~AU . The lower limit
was set by the consideration that the Galactic tide is negligible for distances
less than this, and the upper by the fact that comets with such
large semimajor axes have very short dynamical lifetimes. A starting
eccentricity distribution $n(e)\propto e$ was adopted. The flux of
comets entering a heliocentric sphere of radius 2,000 astronomical
units was computed. Fig.~1 shows the resulting evolution of flux
with time: equilibrium is attained on a timescale of order 300~Myr.

(iii) Analytic formulae due to Fouchard (2004) yield the discrete
change in orbital elements over a single orbit due to the vertical
tide. These were applied to 25,000 comets, and the flux entering
within heliocentric spheres of radii 16 and 8~AU measured. In this
case the initial eccentricity distribution $n(e)$ was taken to be
uniform. Fig.~2 shows the results: essentially the same relaxation
time as before -- 300~Myr -- was found, even although very different
heliocentric spheres and starting $n(e)$ had been adopted.

The semi-analytical and first-order integrations were some orders of
magnitude faster than the direct numerical integrations. They
involve some approximations, but comparison with the numerical
integrations revealed no significant difference in the results
obtained. Synthetic Oort clouds with these relaxed properties were
taken as the starting points for discussing perturbations by a
variable Galactic tide and passing nebulae.

\section {Flux modulation due to the Sun's vertical Galactic motion}

Because of the Sun's vertical motion through the Galactic disc, the
vertical Galactic tide experienced by Oort cloud comets varies
cyclically. The vertical density distribution $\rho(z)$ of ambient
stellar and interstellar material in the disc was taken to decline
exponentially with scale height 60~pc (Joshi 2007), and the vertical
motion of the sun was then given by solving
\begin{equation}
\ddot{z} = -4\pi G \rho(z)z
\end{equation}
The period and amplitude of the solar orbit may then be found, for a
prescribed vertical velocity $v_o$ at $z=0$~pc. The run of density
against time may then be combined with the local tidal force $T$
acting on an Oort cloud comet:
\begin{equation}
T = 4 \pi G \rho(z) z_c
\end{equation}
where $z_c$ represents the vertical height of the comet above or
below the sun.

The Hipparcos data have been used to yield a local in-plane density
$\rho_0 = 0.105~M_\odot\,\rm pc^{-3}$ (Holmberg \& Flynn 2004), but
this appears to be an underestimate by $\sim$50\% due to
incompleteness of discovery of low-luminosity stars
(Garciía-S\'{a}nchez et al. 2001). Stothers (1998) has argued for a
local value $\sim$0.15$\pm0.01~M_\odot\,\rm pc^{-3}$, while Svensmark
(2007) has argued for a mean in-plane density
$\sim0.145\pm0.1~M_\odot\,\rm pc^{-3}$ averaged over the Sun's
traversal of the Galaxy over the last 200~Myr. Adopting the latter
value and assuming $v_0 = 9~\rm km\,s^{-1}$, the variation of flux
of long-period comets into the solar system is derived as shown in
Fig.~3. The tidal cycle has amplitude 2:1, and the peaks are quite
sharp. In Section~5 we argue that this is detectable in the record
of impact cratering on Earth. Depending on the uncertain scale
height of the low-luminosity stars, amplitude ratios of around 2:1
to 5:1 are obtained in these models.

\begin{figure}
\vspace {150pt}
\caption{Top panel: vertical motion of sun in Galactic plane with
local plane density 0.15~$\rm M_\odot\,pc^{-3}$, scale height taken
from the Hipparcos data and solar velocity crossing the plane $v_0
=9~\rm km\,s^{-1}$. Bottom panel: the corresponding variation in
flux of long-period comets into the solar system.}
\label{wicknapfig3}
\end{figure}

\section {Flux modulation due to encounters with nebulae}

In addition to the cyclic component of the long-period comet flux,
sporadic surges are expected due to close encounters with stars and
nebulae. The former have been studied by a number of authors. The
Oort cloud is known to be unstable in the Galactic environment due
to the disruptive effect of encounters with massive nebulae. The
half-life due to such encounters is $\sim$1.9~Gyr (Bailey et al.
1990) and it is generally assumed, although it is not proven, that a
dense inner cloud of comets is available to replenish the loss of
long-period comets.

\begin{figure}
\vspace {150pt}
\caption{Mean interval between encounters with nebulae for various
asymptotic approach speeds, for impact parameter 20~pc, illustrating
the effect of gravitational focusing.} \label{wicknapfig4}
\end{figure}

\begin{figure}
\vspace{150pt}
\caption{Flux of comets entering the planetary system, taken as a
sphere of radius 40~AU, due to a grazing encounter with a GMC. The
flux is computed by direct numerical integration of the cometary
orbits. The GMC has $M = 5\times 10^5~M_\odot$, $p$=20~pc, $V = 15~
\rm km\,s^{-1}$, perihelion occurring at 20~Myr. The initial Oort
Cloud in this simulation comprises 90,000 comets distributed as
$\gamma$=-2 over the range 10,000$\leq r \leq$60,000~AU.}
\label{wicknapfig5}
\end{figure}

Close encounters with, or penetrations of, cold dense nebulae have
occurred quite frequently over geological timescales. Neglecting
gravitational focusing, the solar system passes within $d$~pc of
dark cloud complexes at intervals $\Delta t$~Myr given by
\begin{equation}
\Delta t \sim 800 (M/5\times 10^5)^{0.75} (d/20)^{-2}
\end{equation}
With gravitational focusing, the effective interval is reduced by a
factor
\begin{equation}
\sigma = 1+(V_e/V)^2
\end{equation}
where $V$ represents the asymptotic encounter velocity and $V_e$ is
the escape velocity at the point of closest approach. $\sigma$ is
significant for a close encounter with a massive nebula. For
example, a GMC of mass $M = 5\times 10^5\,M_\odot$ and radius 20~pc
has a surface escape velocity $V_e \sim$15\,$\rm km\,s^{-1}$, and so
for an asymptotic approach speed 15\,$\rm km\,s^{-1}$ the mean
interval between grazing encounters is halved from 800~Myr to
400~Myr, implying about ten such encounters over the period in which
life has existed on Earth. If the density of molecular clouds has
halved over the lifetime of the solar system (Talbot \& Newman
1977), then the expected number of encounters should be increased by
about 25\% over this period. The differential mass distribution of 
molecular clouds is found to be a power law over at least eight 
decades of mass, with index $\alpha = 1.6\pm0.2$ (Mundy 1994). 
From this, the intervals between encounters with nebulae of various 
masses can be found; these are illustrated in Fig.~4 for impact
parameter 20~pc.  It can be seen that, over the Phanerozoic period
(to 600~Myr BP), there has probably been at least one grazing
encounter with a giant molecular cloud, and three or four encounters
with nebulae of mass at least 1.5--2$\times 10^5\,M_\odot$.

Such encounters enhance the comet flux into the planetary system by
filling the loss cones which the Galactic tide cannot reach. In the
present study penetrating encounters were neglected (see Mazeeva
2004 for an analysis of their disruptive effects). Two independent
approaches were again used: direct numerical integration, and
analytic formulae.

For the latter, the trajectory of the molecular cloud was split into
a series of short segments of length $\Delta x$ and the change in
orbital elements arising from the mini-impulse from each segment was
computed. For the latter, formulae given by Roy (1978) were used, in
which $\{\Delta a, \Delta e, \Delta i \}$ etc are given in terms of
a velocity impulse $\Delta v = (\Delta v_S, \Delta v_T, \Delta v_W
)$, where $(\Delta v_S, \Delta v_T)$ are the components of the
impulse in the orbital plane in the radial and transverse
directions, and $\Delta v_W $ is the component normal to the orbit.
The instantaneous acceleration of the comet relative to the sun, due
to each mini-impulse, is given by
\begin {equation}
GM/R^3 (\textbf{R+r}) - GM/R^3 \textbf{R}
\end {equation}
where $\textbf{R}$ represents the instantaneous distance between sun
and molecular cloud, and $\textbf{r}$ is the position vector of the
comet relative to the Sun. Therefore the instantaneous acceleration
of the comet relative to the Sun is
\begin {equation}
GM/R^3 \textbf{r}
\end {equation}
and from an element of trajectory $\Delta x$ along the velocity
vector $V$ of the molecular cloud, the impulsive velocity delivered
to the comet is
\begin {equation}
(\Delta x/V) (GM/R^3) \textbf{r}
\end {equation}
which is simply the vector $\Delta v_S$ in the direction of the
radial coordinate of the comet in its orbit around the Sun. The two
remaining orthogonal velocity components, $\Delta v_T$ and $\Delta
v_W$, conveniently vanish, simplifying the application of the
impulse equations. The evolution of each orbit was computed by the
addition of 40 such mini-impulses. A comparison between this
approach and direct numerical integration yielded essentially
identical results, but the semi-analytic method was orders of
magnitude faster.

Fig.~5 shows the flux of comets into the planetary system due to a
grazing encounter with a giant molecular cloud of mass 5$\times
10^5\,M_\odot$. The comets in this 90,000-particle simulation
initially had a random, isotropic distribution of orbits such that
$n(r)\propto r^{-4}$ in the range 10,000$\leq r\leq$60,000~AU. Their
evolution was followed by numerical integration. The model imposed a 
survival probability of 0.5 on a comet entering the planetary system
during each return, due not only to physical destruction but also to
the risk of ejection into interstellar space through encounters with
the giant planets.  The Figure illustrates the flux of comets entering
the planetary system, represented by a sphere of heliocentric radius
40~AU. We see a distinct bombardment episode, declining with a half
width $\sim$3~Myr. A key feature which emerges is that the
bombardment rate is enhanced while the nebula is still in the
neighborhood of the solar system, and indeed increases for several
million years while the nebula is approaching.

The amplitude of such episodes, relative to the mean background,  depends on the unknown radial
structure of the Oort cloud and is enhanced somewhat for Oort cloud
models with greater central condensation: for example when $\gamma =
-3$, a close encounter with a GMC yields a bombardment episode with
amplitude $A\!\sim$30. Fig.~6 shows the effect of a close encounter
(10~pc) with a 50,000~$M_\odot$ nebula, this time computed with the
semi-analytic approach. Again, a strong bombardment episode is seen,
but its duration is shorter. It appears that the cometary component
of the impact cratering record should be quite `bumpy',
characterised by a periodic component due to the variable Galactic
tide, on which are superimposed discrete surges due to passing
nebulae, an order of magnitude above background.

\begin{figure}
\vspace{150pt}
\caption{Enhanced comet flux due to an encounter with a
50,000~$M_\odot$ nebula at 10~pc; computation is by semi-analytic
formulae as discussed in the text.} \label{wicknapfig6}
\end{figure}

\begin{figure}
\vspace{250pt}
\caption{Top: period/phase distribution of well-dated impact craters
(ages $t\le$250~Myr and dating errors $\le 5$~Myr), obtained by
bootstrap analysis as described in the text. Below: the same for a
synthetic dataset, obtained from a model of the Sun's vertical
motion in the Galaxy coupled with systematic tidal disturbance of
the Oort cloud.} \label{wicknapfig7}
\end{figure}

\section {The impact record}

The record of large impact craters on the Earth appears to show
evidence of a $\sim$36--38~Myr periodicity (Yabushita 2004; Stothers
2006; Napier 2006) and a tendency towards bunching, or occurrence in
impact episodes (Napier 2006). The latter is particularly
conspicuous for impact craters greater than 40 km in diameter.
Breakup of main belt asteroids is inadequate to reproduce this
pattern (loc. cit.), and the question arises whether these
characteristics are quantitatively compatible with Galactic
perturbations of the Oort cloud, yielding impacts either directly
through long-period comet impacts or the intermediary of the
Halley-type comet population.

To examine this further, theoretical cratering flux distributions
were derived from numerical models of the Sun's vertical motion in
the Galactic disc. The impact cratering flux was taken to vary pro
rata with the ambient density (cf Matese et al. 1995), and the
apparent decline of cratering with time (an artefact of the
discovery process) was included. Synthetic data sets were obtained
from this theoretical flux distribution by extracting `craters' from
it at random, in batches of 40 to match the real sample of
high-precision craters (Napier 2006). For the model illustrated in
Fig.~3, $v_0 = 9~\rm km\,s^{-1}$ and the Sun traverses the Galactic
plane every $\sim$39~Myr. In this case the flux varies with
amplitude $\sim$2.

Bootstrap analysis (1000 trials) was then applied to both real and
synthetic datasets. The procedure was to extract data (in sets of
40) at random from the data -- real or artificial -- apply a power
spectrum analysis to each set of 40, and record the peak of the
spectrum wherever it occurred in the range 20--60~Myr. Fig.~7a
illustrates the outcome of one such 1000-trial analysis. It can be
seen that the inbuilt periodicity -- $\sim$39~Myr in this model --
is quite well retrieved, but that some solutions, depending on the
vagaries of data selection, yield harmonics. It thus seems that the
tidal model is broadly able to reproduce the observed periodicity in
the impact cratering record (Fig.~7b), although which signal is the
underlying periodicity and which are harmonics is open to
discussion. These trials indicate that the larger impactors (say
$>$2~km in diameter) are more likely to be comets than the main belt
asteroids, and that the observed bombardment episodes are most
likely of Galactic provenance.

Both our position relative to the Galactic plane (Joshi 2007 and
references therein) and the impact cratering record indicate that we
are presently in, or very close to, the peak of an impact episode.

\section {Discussion and conclusions}

It appears from this numerical study that:

(i) The ebb and flow of the vertical Galactic tide may explain the
quasi-periodic bombardment episodes observed in the record of impact
craters $\ga$40~km in diameter (Napier 2004). There is no known
mechanism capable of yielding similar periodic disturbances of the
asteroid belt. Thus on this model the major impactors largely arrive
in relatively short-lived periodic surges from the Oort cloud.

(ii) To the extent that such episodes generate ecological and
geological disturbances, then the Galactic environment is exerting
control over terrestrial phenomena (Napier \& Clube 1979).

(iii) Order of magnitude enhancements in the comet impact rate occur
during close encounters with massive nebulae and so provide a
crucial link in the chain between ejection of microbiota from the
Earth and their insertion into passing star-forming regions (Napier
2006).

Giant molecular clouds are concentrated within the spiral arms of
the Galaxy. Leitch \& Vasisht (1998) identify two great mass
extinctions, the Cretaceous-Tertiary (65~Myr BP) and end Permian
(225~Myr BP) with Sagittarius-Carina and Scutum-Crux arm crossings
respectively. Gies \& Heisel (2005), on the other hand, find the
mid-points of recent spiral arm crossings at 80 and 156~Myr BP.
Svensmark (2006) has modelled the motion of the Sun in relation to
the spiral arm pattern using a model-dependent hypothesis which has
the Earth's past temperature as a proxy for encounters with spiral
arms. With this model, the solar system passed through the
Sagittarius-Carina arm $\sim$34~Myr BP and the Scutum-Crux arm
$\sim$142~Myr BP. Both these dates coincide with exceptionally
strong bombardment episodes (Napier 2006). It seems that, at
present, uncertainties in both the modelling of spiral arm
kinematics and the strong incompleteness of the impact crater record
preclude a secure identification of impact episodes or mass
extinctions with specific spiral arm crossings.

Both the impact cratering record and the Sun's position near the
Galactic plane imply that we are in a bombardment episode now.

\label{lastpage}

\clearpage

\setcounter{figure}{0}

\begin{figure*}
\includegraphics{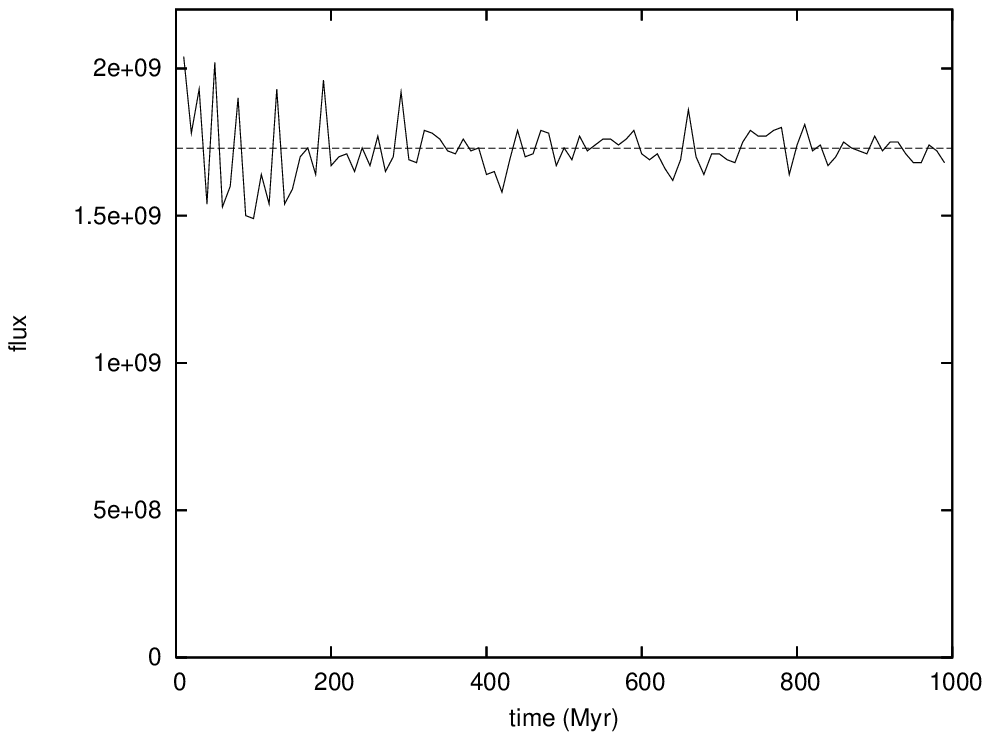}
 \caption{Flux evolution for 1000~Myr under the
influence of the vertical Galactic tide, computed by numerical
integration of formulae due to Klacka \& Gajdosik (1994). The flux 
(arbitrary units) is into a heliocentric sphere of radius 2000~AU, 
and is computed for 50,000 Oort cloud comets with 
$n(a)\propto a^{-2}$ in the range 15,000$\le a\le 30,000$. 
Equilibrium is reached on a timescale of order 300~Myr.}
\end{figure*}
\label{wicknapfig1}

\clearpage

\begin{figure*}
\includegraphics{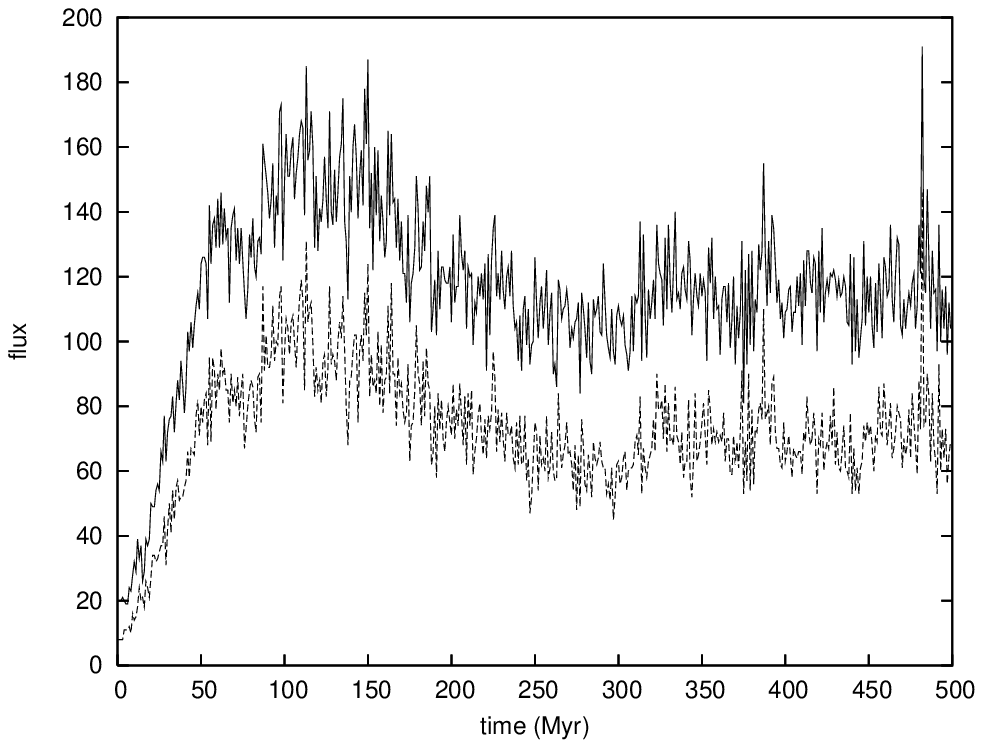}
\caption{Flux evolution for 500~Myr into heliocentric spheres of
radii 16~AU (upper curve) and 8~AU (lower curve) under the influence
of the vertical Galactic tide. 25,000 Oort cloud comets with $n(a)$
as before and a flat initial eccentricity distribution. Computed
from formulae by Fouchard (2005). Equilibrium is again reached on a
timescale of order 300~Myr.}
\end{figure*}
\label{wicknapfig2}

\clearpage

\begin{figure*}
\includegraphics[angle=0,width=0.7\linewidth]{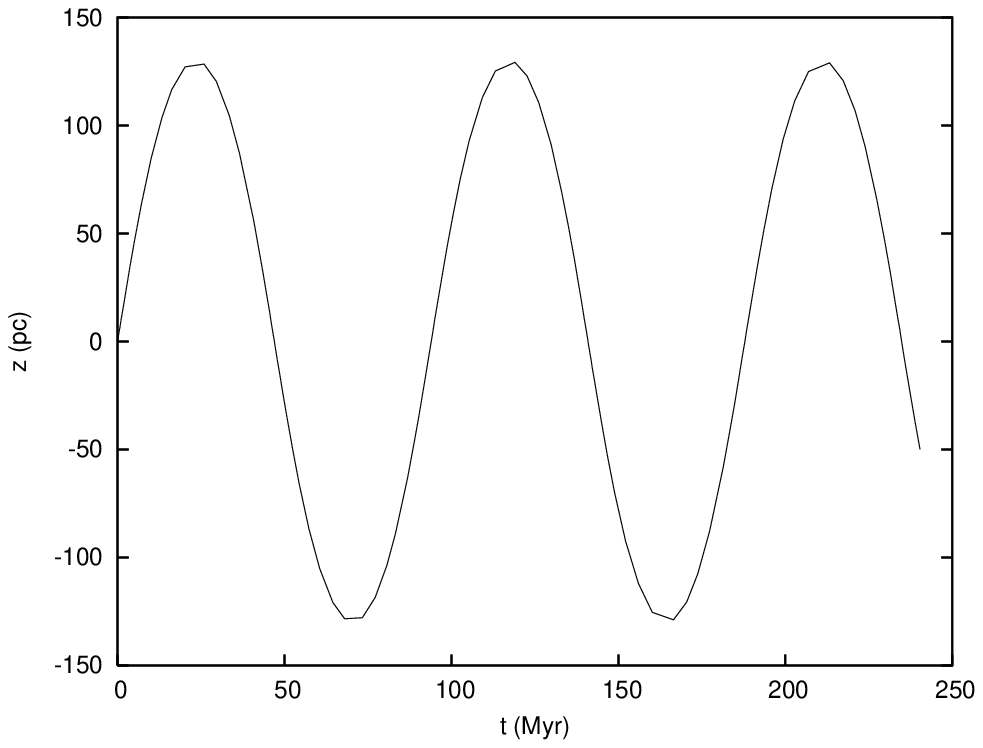}
\includegraphics[angle=0,width=0.7\linewidth]{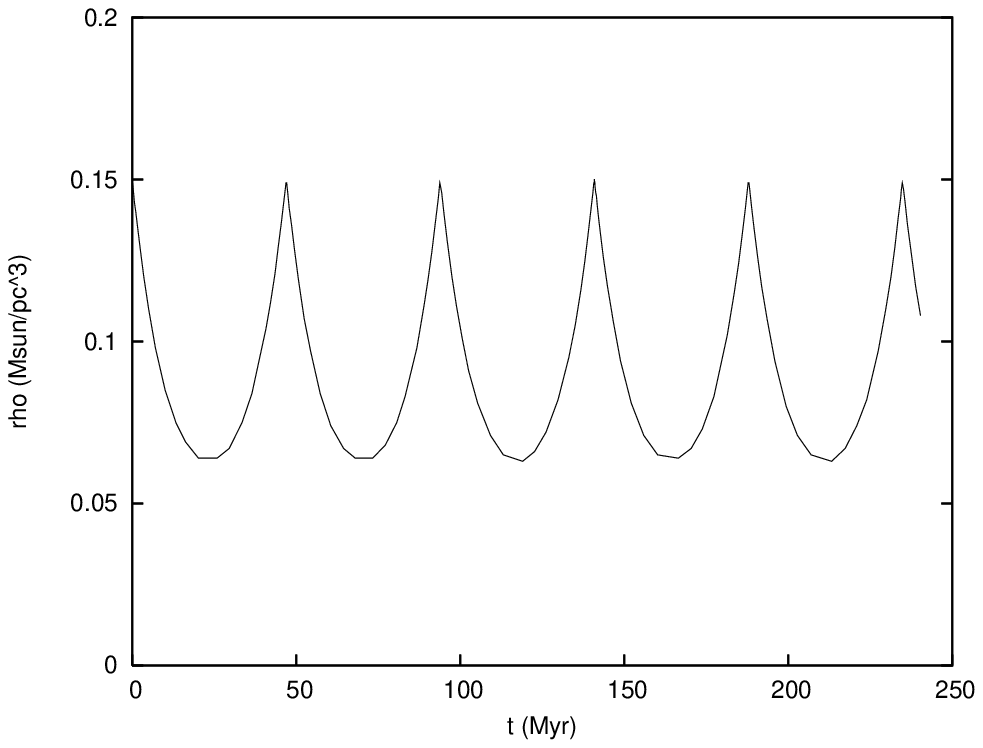}
\caption{Top panel: vertical motion of sun in Galactic plane with
local plane density 0.15~$\rm M_\odot\,pc^{-3}$, scale height taken
from the Hipparcos data and solar velocity crossing the plane $v_0
=9~\rm km\,s^{-1}$. Bottom panel: the corresponding variation in 
local density (and hence, pro rata, the flux of long-period comets 
into the solar system.)}
\label{wicknapfig3}
\end{figure*}

\clearpage

\begin{figure*}
\includegraphics[angle=0,width=\linewidth]{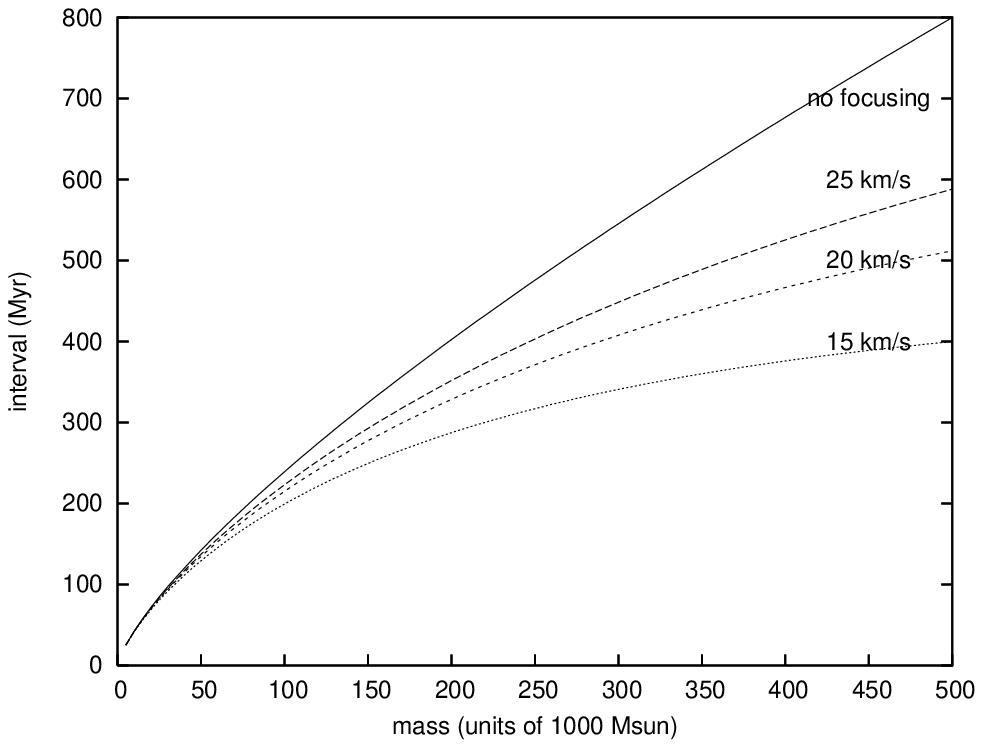}
\caption{Mean interval between encounters with nebulae for various
asymptotic approach speeds, for impact parameter 20~pc, illustrating
the effect of gravitational focusing.} \label{wicknapfig4}
\end{figure*}

\clearpage

\begin{figure*}
\includegraphics[angle=0,width=0.9\linewidth]{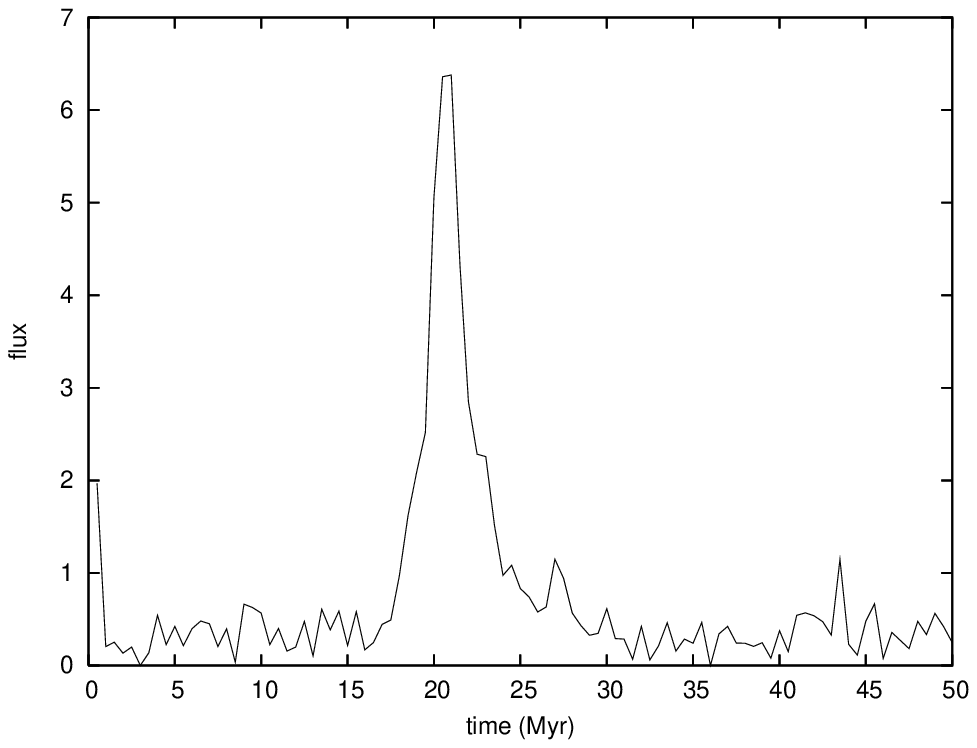}
\caption{Flux of comets entering the planetary system, taken as a
sphere of radius 40~AU, due to a grazing encounter with a GMC. The
flux is computed by direct numerical integration of the cometary
orbits. The GMC has $M = 5\times 10^5~M_\odot$, $p$=20~pc, $V = 15~
\rm km\,s^{-1}$, perihelion occurring at 20~Myr. The initial Oort
Cloud in this simulation comprises 90,000 comets distributed as
$\gamma$=-2 over the range 10,000$\leq r \leq$60,000~AU.}
\label{wicknapfig5}
\end{figure*}

\clearpage

\begin{figure*}
\includegraphics[width=0.8\linewidth]{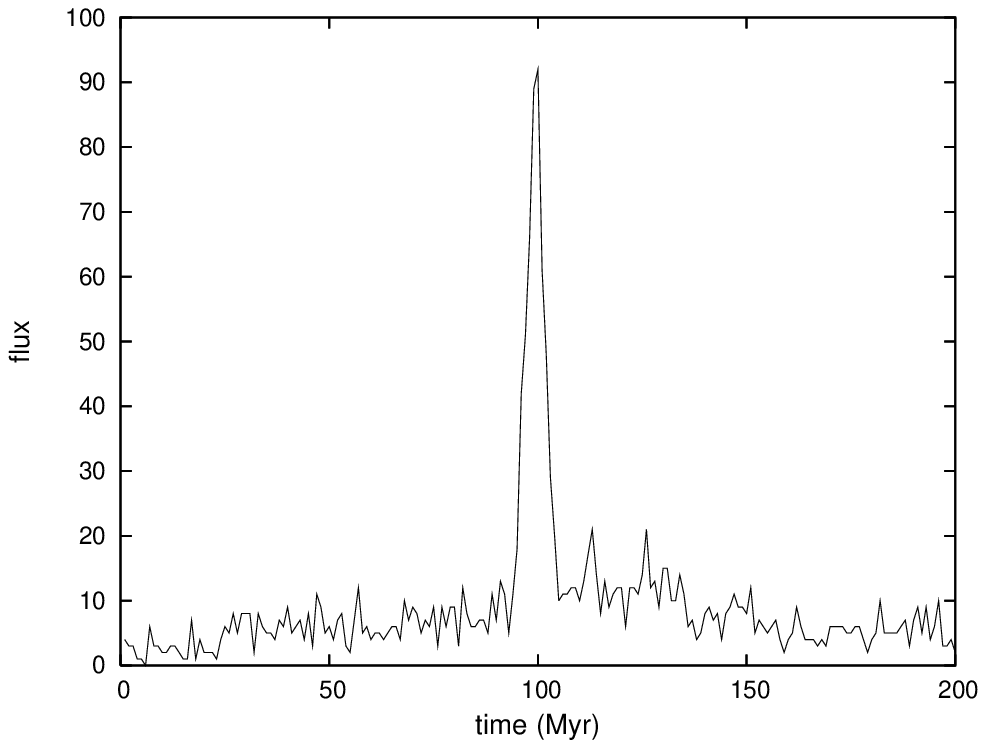}
\caption{Enhanced comet flux due to an encounter with a
50,000~$M_\odot$ nebula at 10~pc; computation is by semi-analytic
formulae as discussed in the text.} \label{wicknapfig6}
\end{figure*}

\clearpage

\begin{figure*}
\includegraphics[angle=270,width=0.6\linewidth]{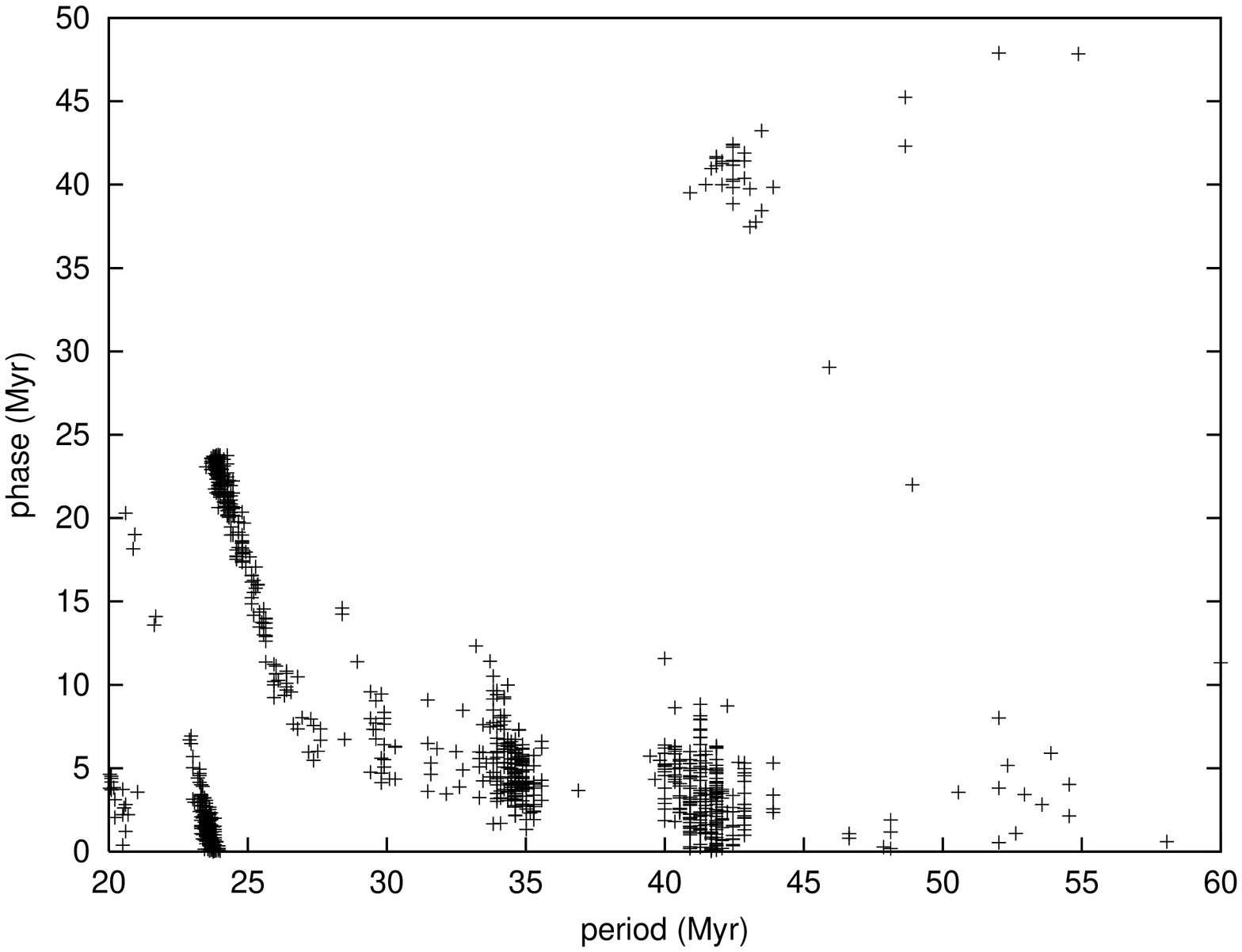}
\includegraphics{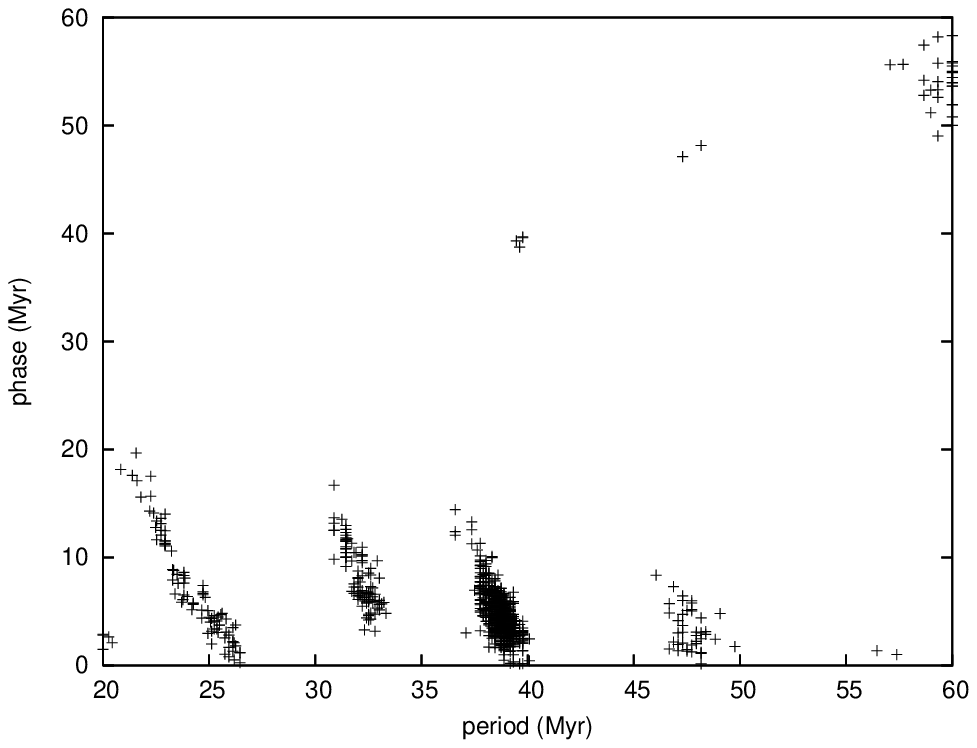}
\caption{Top: period/phase distribution of well-dated impact craters
(ages $t\le$250~Myr and dating errors $\le 5$~Myr), obtained by
bootstrap analysis as described in the text. Below: the same for a
synthetic dataset, obtained from a model of the Sun's vertical
motion in the Galaxy coupled with systematic tidal disturbance of
the Oort cloud.} \label{wicknapfig7}
\end{figure*}

\end{document}